\documentclass[12pt]{article}
\usepackage{cite,amssymb,amsmath,hyperref,authblk}
\begin{document}
\title{\bf Gauge Fields in the 5D Gravity-Scalar Standing Wave Braneworld}
\author[1,2]{Merab Gogberashvili \thanks{gogber@gmail.com}}
\author[3]{Pavle Midodashvili \thanks{pmidodashvili@yahoo.com}}
\affil[1]{Andronikashvili Institute of Physics, 6 Tamarashvili St., Tbilisi 0177, Georgia}
\affil[2] {Javakhishvili State University, 3 Chavchavadze Ave., Tbilisi 0179, Georgia}
\affil[3]{Ilia State University, 3/5 Cholokashvili Ave., Tbilisi 0162, Georgia}
\date{}
\maketitle
\begin{abstract}
We investigate localization problem for gauge fields within the 5D standing wave braneworld with real scalar field and show that there exist normalizable vector field zero modes on the brane.
\end{abstract}

\vskip 0.5cm

Most of the braneworld scenarios considered in literature are realized using time-independent field configurations. However, also some models with time-dependent metrics and matter fields were proposed \cite{S}. Localization problem is the key issue in realization of the braneworld scenario. For reasons of economy and stability one would like to have a universal gravitational trapping mechanism for all fields. In brane models the implementation of a pure gravitational trapping mechanism for vector fields remains most problematic \cite{vect}.

Recently we have considered the new class of the non-stationary braneworlds \cite{Wave}, which provides the natural localization mechanism by the collective oscillations of the bulk gravitational and scalar fields \cite{Loc}. In this letter we investigate the vector field localization problem within the 5D standing wave braneworld with a real scalar field introduced in \cite{RealScalarField}.

The metric of the standing wave braneworld \cite{RealScalarField} has the form:
\begin{equation} \label{Metric}
 ds^2 = \frac {e^S}{(1 - k|r|)^{2/3}}\left( dt^2 - dr^2 \right) - (1 - k|r|)^{2/3}\left( e^u dx^2 + e^u dy^2 + e^{-2u}dz^2 \right)~,
\end{equation}
where $k$ is a positive constant. The metric functions $u$ and $S$ (the solutions to the 5D Einstein equations) are:
\begin{eqnarray} \label{MetricFunctions}
 u(t,|r|)&=& A \sin (\omega t) J_0 \left(\frac{\omega }{k} - \omega|r|\right)~,\nonumber\\
 S(|r|) &=& \frac{3\omega ^2(1 - k|r|)^2}{2k^2}A^2\left[ J_0^2\left( \frac{\omega }{k} - \omega |r| \right) + J_1^2\left( \frac{\omega }{k} - \omega |r| \right) - \right. \\
 &&\left. - \frac{k}{\omega (1 - k|r|)} J_0\left( \frac{\omega }{k} - \omega |r| \right)J_1\left( \frac{\omega }{k} - \omega |r| \right) - \frac{1}{(1 - k|r|)^2}J_1^2\left( \frac{\omega }{k} \right) \right]~,\nonumber
\end{eqnarray}
where $A$ and $\omega$ are the amplitude and frequency of standing waves, and $J_0$ and $J_1$ are Bessel functions of the first kind.

We have also the solution to the 5D Klein-Gordon equation in the background metric (\ref{Metric}),
\begin{equation} \label{BackgroundBulkScalarField}
  \varphi(t,|r|) = \sqrt {\frac{3 M^3}{2}} ~A \cos (\omega t) J_0\left(\frac{\omega }{k} - \omega|r|\right)
\end{equation}
($M$ denotes the 5D fundamental scale), describing the bulk standing waves of the real scalar field.

The metric (\ref{Metric}) has the horizon at $|r| = 1/k$, where some components of the Ricci tensor get infinite values, while all gravitational invariants are finite. This resembles the situation with the Schwarzschild Black Hole, however, the determinant of (\ref{Metric}),
\begin{equation}\label{MetricDeterminant}
  \sqrt g  = (1 - k|r|)^{1/3}e^S~,
\end{equation}
becomes zero at the horizon. As a result nothing can cross this horizon and matter fields are confined within the distance $\sim 1/k$ in the extra space. If the curvature scale, $k$, is relatively small, the horizon size is large and the width of the brane is determined by the metric function $S (|r|)$ \cite{RealScalarField}. So trapping of matter fields in this case is caused by the 'pressure' of the bulk oscillations and not by the existence of the horizon in the extra space.

The solutions (\ref{MetricFunctions}) and (\ref{BackgroundBulkScalarField}) describe the scalar and gravitational standing waves in the extra dimension, $r$, which are bounded by the brane at $r=0$ and the horizon at $|r| = 1/k$. The metric function, $u(t,|r|)$, and the scalar field, $\varphi(t,|r|)$, are oscillating $\pi/2$ out of phase in time, i.e. the energy of the oscillations passes back and forth between the gravitational and scalar field standing waves. On the other hand, these functions have a similar dependence on the spatial coordinate $r$. Their $r$-dependent factors vanish at the points where the Bessel function $J_0 (\omega /k - \omega|r|)$ is zero, and form the nodes of the bulk standing waves. One of the nodes is located at the origin, $r=0$, where (due to the presence of absolute value of $r$ in (\ref{Metric})) the Ricci tensor has $\delta$-like singularity that is smoothed by the brane. The bulk oscillations of physical quantities should vanish also at the horizon, $|r| = 1/k$, that actually determines the size of the extra space for the brane observer.

Now let us show existence of the pure gravitational localization of vector field zero modes on the brane by the background metric (\ref{Metric}). We investigate only the $U(1)$ vector fields, the generalization to the non-Abelian case is straightforward. The standard action of the 5D massless vector field $A_M$ is:
\begin{equation}\label{ActionOfVectorField}
  S =  - \frac 14 \int d^5x\sqrt g~ g^{BD}g^{MN}F_{BM}F_{DN}~ ,
\end{equation}
where the 5D vector field tensor is defined as:
\begin{equation}\label{VectorFieldTensor}
  F_{BM} = \partial_B A_M - \partial_M A_B~.
\end{equation}
From the action (\ref{ActionOfVectorField}) we get the 5D Maxwell equations,
\begin{equation}\label{VectorFieldEquations}
  \frac{1}{\sqrt g }~\partial_B \left( \sqrt g g^{BD} g^{MN} F_{DN} \right) = 0~.
\end{equation}

In the vicinity of standing wave nodes the oscillations of the metric functions in (\ref{Metric}) are small and when $|r| \to 0$ a solution to (\ref{VectorFieldEquations}) can be factorized in the form:
\begin{eqnarray}\label{VectorFieldAnsatz}
  A_t\left(x^C\right) &=& (1 - k|r|)^{-2/3}e^{S(r)}   \xi (|r|) a_t\left(x^\mu\right)~, \nonumber \\
  A_i\left(x^C\right) &=& (1 - k|r|)^{2/3}e^{u(t,r)}  \rho(|r|) a_i\left(x^\mu\right)~, \nonumber ~~~~~~(i = x, y) \\
  A_z\left(x^C\right) &=& (1 - k|r|)^{2/3}e^{-2u(t,r)}\rho(|r|) a_z\left(x^\mu\right)~, \\
  A_r\left(x^C\right) &=& 0~, \nonumber
\end{eqnarray}
where $a_\mu \left(x^\nu\right)$ are the components of the vector potential on the brane (Greek letters are used for 4D indices and $i$ runs over $x$ and $y$). The last equation in (\ref{VectorFieldAnsatz}) in fact is our 5D gauge condition.

Using the factorization (\ref{VectorFieldAnsatz}) the components of the 5D tensor (\ref{VectorFieldTensor}) can be explicitly written as:
\begin{eqnarray}\label{5DVectorFieldComponents}
  F_t^i \left(x^A \right)&=& \rho f_t^i + \rho a^i \partial _tu + \left[ \rho  - \frac{e^{S - u}}{( 1 - k|r|)^{4/3}}\xi \right]\partial ^i a_t~, \nonumber \\
  F_t^z \left(x^A \right)&=& \rho f_t^z - 2\rho a^z\partial _tu + \left[ \rho  - \frac{e^{S - u}}{( {1 - k|r|)^{4/3}}\xi } \right]\partial ^za_t~, \nonumber \\
  F_x^y \left(x^A \right)&=& \rho f_x^y~, ~~~~~ F_i^z \left(x^A \right) = \rho f_i^z + \rho \left[1 - e^{3u} \right]\partial ^z a_i~,\\
  F_r^t \left(x^A \right)&=& \left[ \xi ' + \xi \left( S' + \frac 23 \cdot \frac{k~ \mathrm{sgn}(r)}{1 - k|r|} \right) \right]a^t~, \nonumber\\
  F_r^i \left(x^A \right)&=& \left[\rho ' + \rho \left(u' - \frac 23 \cdot \frac{k~ \mathrm{sgn}(r)}{1 - k|r|} \right) \right]a^i~, \nonumber \\
  F_r^z \left(x^A \right)&=& \left[\rho ' - \rho\left(2u' + \frac 23 \cdot \frac{k~ \mathrm{sgn}(r)}{1 - k|r|} \right) \right]a^z \nonumber
\end{eqnarray}
(where $i = x,y$), and components of 4D field on the brane are:
\begin{equation}\label{4DVectorFieldComponents}
  f_\mu ^\nu \left(x^\alpha \right) = \partial _\mu a^\nu  - \partial ^\nu a_\mu ~.
\end{equation}

Using (\ref{5DVectorFieldComponents}) we find the components of the 5D Maxwell equations (\ref{VectorFieldEquations}):
\begin{eqnarray}\label{VectorFieldEquationsT}
  (1 - k|r|)^{-1}\partial_r\left\{ (1 - k|r|)\left[ \xi' + \xi \left( S' + \frac 23 \cdot \frac{k~\mathrm{sgn}(r)}{1 - k|r|} \right) \right] \right\}a_t - \\
  - \left[ \partial_t^2 - ( 1 - k|r|)^{-4/3}e^{S-u}\left( \partial_i\partial^i + e^{3u}\partial_z^2 \right) \right]\xi a_t +  \nonumber\\
  + \partial_t\left[ \xi \partial_ta_t - \rho \left( \partial_ia^i + \partial_za_z \right)\right] - \dot u \rho \left[ \partial_i a^i - 2\partial_z a_z \right] = 0~,\nonumber
\end{eqnarray}
\begin{eqnarray}\label{VectorFieldEquationsX}
  (1 - k|r|)^{-1}\partial_r\left\{ (1 - k|r|)\left[ \rho' + \rho \left( u' - \frac 23 \cdot \frac{k~\mathrm{sgn}(r)}{1 - k|r|} \right)\right] \right\}a_i - \\
  - \left[ \partial_t^2 - ( 1 - k|r|)^{-4/3}e^{S-u}\left( \partial_i\partial^i + e^{3u}\partial_z^2 \right) \right]\rho a_i +  \nonumber\\
  + (1 - k|r|)^{-4/3}e^{S-u} \partial_i \left[ \xi\partial_ta_t - \rho \left( \partial_i a^i + \partial_z a_z \right) \right] -  \nonumber\\
  - \left[ \rho \left( \ddot u + \dot u \partial_t \right)a_i + (1 - k|r|)^{-4/3}e^{S-u}\dot u\xi \partial_i a_t \right] = 0~, \nonumber
\end{eqnarray}
\begin{eqnarray}\label{VectorFieldEquationsZ}
  (1 - k|r|)^{-1}\partial_r\left\{ (1 - k|r|) \left[ \rho ' + \rho \left(- 2u' - \frac 23 \cdot \frac{k~\mathrm{sgn}(r)}{1 - k|r|}\right) \right] \right\} a_z - \\
  - \left[ \partial_t^2 - ( 1 - k|r|)^{-4/3}e^{S-u}\left( \partial_i\partial^i + e^{3u}\partial_z^2 \right) \right]\rho a_z +  \nonumber \\
  + (1 - k|r|)^{-4/3}e^{S+2u}\partial _z \left[ \xi\partial_ta_t - \rho \left( \partial_i a^i + \partial_z a_z \right) \right]  + \nonumber\\
  + 2\left[ \rho \left( \ddot u + \dot u \partial_t \right)a_z + (1 - k|r|)^{-4/3}e^{S+2u}\dot u\xi\partial_z a_t \right] = 0~, \nonumber
 \end{eqnarray}
\begin{eqnarray}\label{VectorFieldEquationsR}
  \left[ \xi' + \xi \left( S' + \frac 23 \cdot \frac{k~\mathrm{sgn}(r)}{1 - k|r|} \right) \right]\partial_t a_t - \left[ \rho' + \rho \left( u' - \frac 23 \cdot \frac{k~\mathrm{sgn}(r)}{1 - k|r|} \right) \right]\partial_i a^i -  \\
  - \left[ \rho' + \rho \left(- 2u' - \frac 23 \cdot \frac{k~\mathrm{sgn}(r)}{1 - k|r|}\right) \right]\partial_z a_z = 0~,\nonumber
\end{eqnarray}
where overdots and primes denote derivatives with the respect to $t$ and $|r|$, respectively.

To have a localized field on a brane, 'coupling' constants appearing after integration of the field Lagrangian over the extra coordinate must be non-vanishing and finite. So we require that the extra dimension factors $\xi_0(|r|)$ and $\rho_0(|r|)$ of the vector field zero mode wavefunction have a sharp peak on the brane in the extra space, i.e. on the brane we impose the following boundary conditions on them:
\begin{equation}\label{BoundaryConditionsOnBrane}
  \left. \frac{\xi_0'}{\xi_0 } \right|_{|r| \to 0} \gg S'|_{|r| \to 0}~, ~~~~~\left. \frac{\xi_0'}{\xi_0 } \right|_{|r| \to 0} \gg k~, ~~~~~\left. \frac{\rho_0'}{\rho_0 } \right|_{|r| \to 0} \gg u'|_{|r| \to 0}~, ~~~~~ \left. \frac{\rho_0'}{\rho_0} \right|_{|r| \to 0} \gg k~.
\end{equation}
Then the fifth component of Maxwell equations (\ref{VectorFieldEquationsR}) takes the standard 4D Lorentz gauge condition form,
\begin{equation}\label{4DGaugeCondition}
  \eta ^{\alpha \beta }{\partial _\alpha }{a_\beta } = 0~.
\end{equation}

To show localization of the gauge fields, i.e. the existence of the functions $\xi_0(|r|)$ and $\rho_0(|r|)$ which have maximum on the brane, $|r|\to 0$, and vanish at the horizon, $|r|\to 1/k$,  we investigate the Maxwell equations (\ref{VectorFieldEquationsT}), (\ref{VectorFieldEquationsX}), (\ref{VectorFieldEquationsZ}) and (\ref{VectorFieldEquationsR}) in these two limiting regions.

Bulk standing waves have node on the brane, so, close to the brane, $|r| \to 0$, we can use the separation of the variables in the form (\ref{VectorFieldAnsatz}). We seek for the zero mode solution and assume that on the brane it corresponds to the 4D vector plane wave:
\begin{equation}\label{4DVectorField}
  a_\mu \left(x^\nu\right) \sim \varepsilon_\mu e^{i\left( Et + p_xx + p_yy + p_zz\right)}~,
\end{equation}
where $E$, $p_x$, $p_y$ and $p_z$ are components of the energy-momentum of vector field on the brane. Then the components of the Maxwell equations (\ref{VectorFieldEquationsT}), (\ref{VectorFieldEquationsX}), (\ref{VectorFieldEquationsZ}) and (\ref{VectorFieldEquationsR}) for the vector field zero mode wavefunction (\ref{4DVectorField}) reduce to the simple equations for the extra dimension factors:
\begin{equation}\label{EquationRho}
  \rho_0'' + B|r|\rho_0  = \xi_0'' + B|r|\xi_0 = 0~,
\end{equation}
where the constant $B$ is:
\begin{equation}
B = \frac {3}{2k} E^2 \omega^2 A^2 J_1^2\left(\frac{\omega}{k}\right)~.
\end{equation}
The equations (\ref{EquationRho}) have the following exact solutions:
\begin{equation}\label{}
  \xi_0(|r|) = \rho_0 (|r|) = C_1Ai\left( -\sqrt[3]{B}~|r| \right) + C_2Bi\left( -\sqrt[3]{B}~|r| \right)~,
\end{equation}
where $C_1$ and $C_2$ are integration constants and $Ai$ and $Bi$ denote Airy functions. To fulfill the boundary conditions (\ref{BoundaryConditionsOnBrane}) we set $C_1=0$, and the series expansions of extra dimension factors near the brane will be:
\begin{eqnarray}\label{RhoOnTheBrane}
  \xi_0 (|r|)|_{|r| \to 0} = \rho_0 (|r|)|_{|r| \to 0} = \frac {C_2}{3^{1/6}\Gamma (2/3)}\left[ 1 - \frac{3^{5/6}\Gamma^2 (2/3)}{2\pi} \sqrt[3]{B}~|r| - \frac 16 B|r|^3 \right] + O\left(r^4\right).
\end{eqnarray}
It is easy to see that the boundary conditions (\ref{BoundaryConditionsOnBrane}) can be fulfilled only for the vector field zero modes with the energy:
\begin{equation}
  E \gg  k A^2~,
\end{equation}
where the amplitude of standing waves $A \gg 1$.

In the second limiting region, $|r| \to 1/k$, oscillations are small and we can again use the separation of variables (\ref{VectorFieldAnsatz}). The systrm of equations (\ref{VectorFieldEquationsT}), (\ref{VectorFieldEquationsX}), (\ref{VectorFieldEquationsZ}) and (\ref{VectorFieldEquationsR}) in this region has the solution:
\begin{eqnarray}\label{SolutionAtHorizon}
  \xi_0 (|r|) &=& c_1 (1 - k|r|)^{2/3}~, \nonumber \\
  \rho_0(|r|) &=& c_2 ~, \\
  a{_\mu }\left(x^\nu \right) &=& const~, \nonumber
\end{eqnarray}
where $c_1$ and $c_2$ are some constants.

For the sharply decreasing extra dimension factors $\xi_0 (|r|)$ and $\rho_0 (|r|)$ with the asymptotes (\ref{RhoOnTheBrane}) and (\ref{SolutionAtHorizon}) the 5D vector field action (\ref{ActionOfVectorField}) is integrable over the extra coordinate $r$. This means that the vector field zero mode wavefunction is localized on the brane. Indeed for our factorization (\ref{5DVectorFieldComponents}) the 5D vector field Lagrangian close to the brane, $r \to 0$, reduces to:
\begin{equation}\label{5DVectorFieldLagrangianOnBrane}
  L = - \frac 14\sqrt g F_M^NF_N^M |_{|r|\to 0} \rightarrow - \frac 14 \left[\rho_0^2 (0) f_\mu ^\nu f_\nu ^\mu + \rho_0 '^2 (0)a_\nu a^\nu\right] ~.
\end{equation}
This expression contains the mass terms $ \sim \rho_0 '^2a_\nu a^\nu $, i.e. photons are massless only on the brane and acquire large masses in the bulk. The behavior of this type is similar to the Meissner effect in superconductors and can be considered as another manifestation of our gravitational trapping mechanism.

To conclude, in this paper we have explicitly shown the existence of normalizable zero modes of gauge bosons on the brane within the 5D standing wave braneworld with real scalar field.

\medskip


\noindent {\bf Acknowledgments:} MG was partially supported by the grant of Shota Rustaveli National Science Foundation $\#{\rm DI}/8/6-100/12$. The research of PM was supported by Ilia State University.


\end{document}